\documentclass{article}

\usepackage{microtype}
\usepackage{graphicx}
\usepackage{subcaption}
\usepackage{booktabs} 

\usepackage[hyperfootnotes=false]{hyperref}

\usepackage[accepted]{icml2026}

\usepackage{amsmath}
\usepackage{amssymb}
\usepackage{mathtools}
\usepackage{amsthm}
\usepackage{multirow}
\usepackage{xcolor}
\usepackage{fvextra}
\usepackage{tabularx}
\usepackage{ragged2e}
\usepackage{enumitem} 
\newcolumntype{Y}{>{\RaggedRight\arraybackslash}X}
\newcolumntype{Z}{>{\RaggedRight\arraybackslash}X}
\usepackage[capitalize,noabbrev]{cleveref}

\theoremstyle{plain}

\theoremstyle{definition}

\theoremstyle{remark}

\usepackage[disable, textsize=tiny]{todonotes}

\icmltitlerunning{BaRA: Budget-constrained and Reliable Web Data Collection Agent}

\begin{document}

\twocolumn[
  \icmltitle{BaRA: Budget-constrained and Reliable Web Data Collection Agent}



  \icmlsetsymbol{equal}{*}

  \begin{icmlauthorlist}
    \icmlauthor{Soojeong Lee}{equal,yonsei}
    \icmlauthor{Joseph Lee}{equal,yonsei}
    \icmlauthor{Yongseong Cho}{etri}
    \icmlauthor{Sunjae Kim}{hustlers}
    \icmlauthor{Youngwoo Moon}{hustlers}
    \icmlauthor{Kyungwoo Song}{yonsei}
  \end{icmlauthorlist}

  \icmlaffiliation{yonsei}{
    Department of Statistics and Data Science,
    Yonsei University,
    Seoul,
    Republic of Korea
  }

  \icmlaffiliation{etri}{
    Electronics and Telecommunications Research Institute (ETRI),
    Daejeon,
    Republic of Korea
  }

  \icmlaffiliation{hustlers}{
    HUSTLERS Corp.,
    Seoul,
    Republic of Korea
  }

  \icmlcorrespondingauthor
    {Kyungwoo Song}
    {kyungwoo.song@yonsei.ac.kr}

  \icmlkeywords{Web Automation,Browser Agents,Large Language Models,Planning Systems,Web Navigation,Multimodal Extraction}


  \vskip 0.3in
]



\printAffiliationsAndNotice{\icmlEqualContribution}  

\begin{abstract}
Large language model (LLM)-based web agents automate web navigation and data collection. However, live web data collection demands capabilities beyond task completion: agents must discover site-internal pages and retrieve text, image, and video artifacts in an accessible form within a fixed interaction budget. We formulate this setting as budget-constrained, site-level multimodal web data collection and propose \textbf{B}udget-constrained \textbf{a}nd \textbf{R}eliable \textbf{A}gent (BaRA). BaRA performs breadth-first search (BFS)-based link discovery with liveness verification to filter hallucinated and dead links, then validates extracted multimodal artifacts using rule-based provenance and accessibility checks. A history-based self-reflection module recovers from execution failures and incomplete outputs. On controlled synthetic and real-world websites, BaRA consistently improves valid-link discovery and download-valid multimodal extraction over existing agents. Our code is available at \url{https://github.com/MLAI-Yonsei/BaRA-Agent}. 
\end{abstract}

\begin{figure*}[t]

\centering

\includegraphics[width=\textwidth]{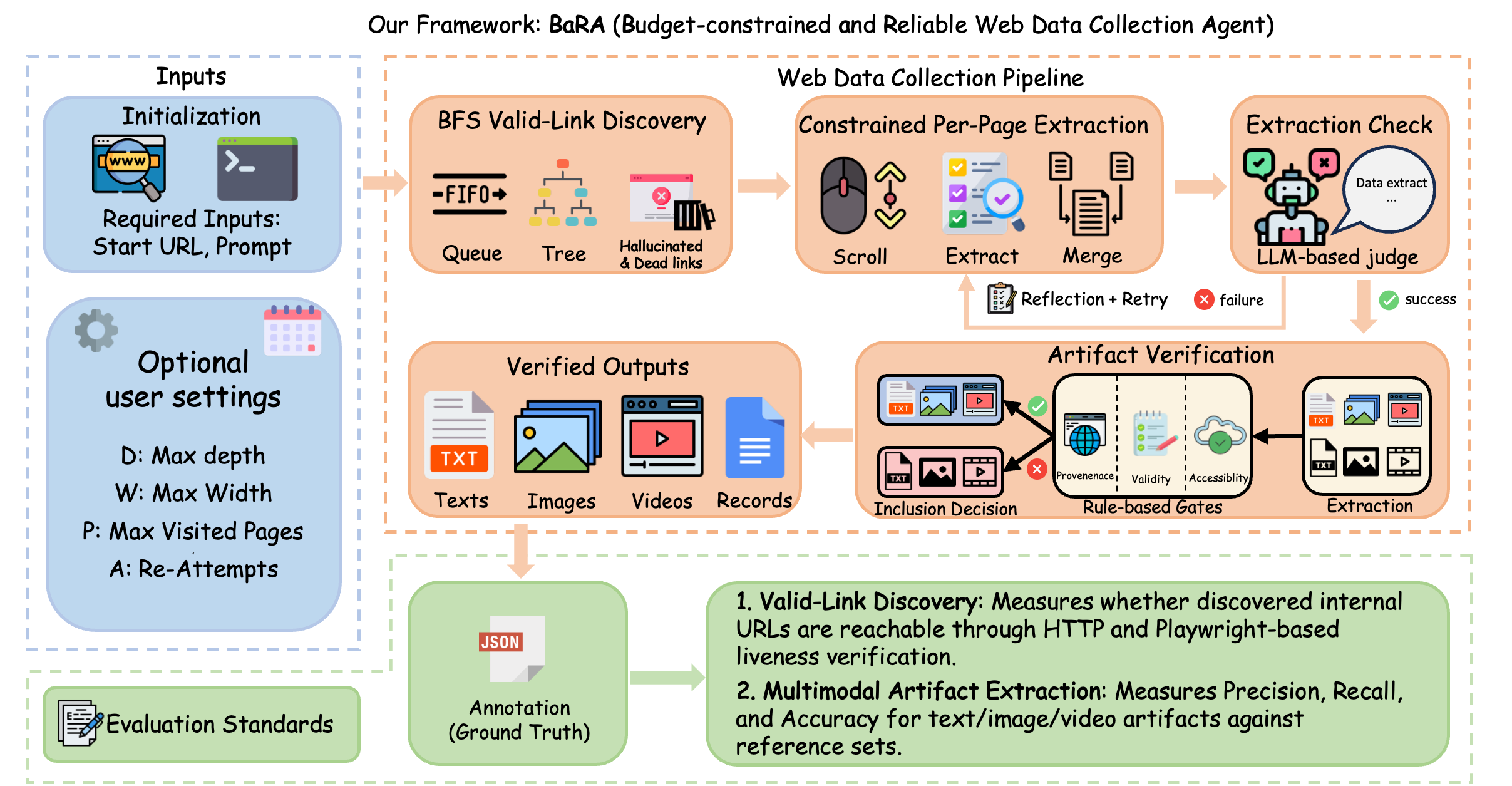}

\caption{Overview of BaRA. Starting from a seed URL and user-configurable budgets (maximum BFS depth $D$, maximum retained child links per page $W$, maximum visited pages $P$, and maximum number of retry attempts $A$), the system first performs bounded BFS traversal, then applies constrained per-page extraction to each discovered page, and finally validates and downloads the recovered media. When exploration or extraction fails, a history-based self-reflection module rewrites the prompt for the next attempt. The pipeline returns download-valid artifacts together with per-page metadata for evaluation.}

\label{fig:overview}

\end{figure*}

\section{Introduction}

Web data collection remains a practical bottleneck for natural language processing (NLP) and multimodal pipelines, including dataset construction, information extraction, and retrieval-augmented generation. LLM-based web agents make interaction with live websites more flexible, but existing web-agent benchmarks mainly evaluate \emph{instruction-conditioned task completion}: whether an agent reaches a target state for a given task. They do not directly target site-internal page coverage or the validity of collected artifacts. In live web data collection, robust crawling and extraction remain fragile, as agents may miss relevant pages, return incomplete multimodal outputs, or produce media URLs that are hallucinated or not downloadable.

We study a distinct setting: \emph{budget-constrained, site-level multimodal web data collection}. Given a seed URL, the goal is to discover site-internal pages within a fixed interaction budget and return provenance-grounded text, image, and video artifacts in an accessible form. Unlike task-completion benchmarks, this setting treats the validity of discovered site-internal links and extracted artifacts as primary evaluation targets. Although prior benchmarks often impose step or time limits, they typically do not measure whether limited interactions are converted into verified multimodal artifacts. To address this gap, we propose BaRA, a framework for reliable live web data collection.

\paragraph{Contributions.}
Our contributions are threefold:
\begin{itemize}
    \item \textbf{Problem formulation.}
    We formulate \emph{budget-constrained, site-level multimodal web data collection} as a distinct evaluation setting for web agents.

    \item \textbf{Method.}
    We propose BaRA, which orchestrates BFS-based link discovery, liveness verification, reflection-guided recovery, and rule-based artifact verification under a fixed interaction budget.

    \item \textbf{Benchmarks and evaluation.}
    We introduce two benchmarks: a 50-site synthetic benchmark with known structure and a 50-site real-world benchmark constructed from Tranco \citep{pochat2018tranco} long-tail domains and Hacker News \citep{algolia_hn_api} under automatic filtering and diversity constraints. Experiments on both benchmarks show that BaRA improves valid-link discovery and download-valid multimodal extraction over existing agents.
\end{itemize}

Figure~\ref{fig:overview} provides an overview of BaRA's end-to-end pipeline, from budgeted BFS valid-link discovery to multimodal extraction and artifact verification.

\section{Related Work}

\label{sec:related}

\paragraph{Web-Agent Benchmarks and Environments.}
Research on web and graphical user interface (GUI) agents has evolved from controlled browser environments such as World of Bits and MiniWoB++ \citep{shi2017world, liu2018reinforcement} to more realistic and large-scale settings. WebShop introduced a simulated e-commerce environment with real products and compositional instructions \citep{yao2022webshop}, while Mind2Web proposed a large-scale benchmark for instruction-conditioned task completion on real websites \citep{deng2023mind2web}. Subsequent benchmarks, including WebArena and VisualWebArena \citep{zhou2023webarena, koh2024visualwebarena}, WorkArena and WorkArena++ \citep{drouin2024workarena, boisvert2024workarena++}, and WebLINX \citep{lu2024weblinx}, further expanded evaluation to autonomous interaction, enterprise knowledge work, and conversational web navigation. Online-Mind2Web further highlights the gap between benchmark performance and robustness in live web environments \citep{xue2025illusion}.

These benchmarks primarily evaluate task completion or target-state achievement for a specified instruction. BaRA targets a different setting: \emph{budget-constrained, site-level multimodal web data collection}. Rather than measuring task completion alone, BaRA evaluates site-internal page discovery under a fixed interaction budget and the return of provenance-grounded text, image, and video artifacts in an accessible form. Accordingly, the core evaluation targets are valid-link discovery and download-valid multimodal extraction, rather than task success rate alone.

\paragraph{Web-Agent Methods and Reflection.}
WebGPT demonstrated improvements in grounded long-form question answering through browser interaction \citep{nakano2021webgpt}. WebVoyager and SeeAct advanced multimodal grounding for interaction with live websites \citep{he2024webvoyager, zheng2024gpt}. In particular, SeeAct uses visual observations of webpages to ground actions to HTML elements, illustrating how LLM-based agents can interact with dynamic page structures. Other work has explored prompting, memory, and agent design choices, including WebGUM \citep{furuta2023multimodal}, Synapse \citep{zheng2023synapse}, and AgentOccam \citep{yang2024agentoccam}, and is closely related to reasoning and self-improvement paradigms such as ReAct and Reflexion \citep{yao2022react, shinn2023reflexion}.

BaRA builds on this line of work but applies reflection to live web data collection rather than instruction-conditioned task completion. Its reflection module uses execution history to recover from failures and incomplete outputs, while its BFS-based link discovery and liveness verification reduce hallucinated and dead links before they waste the fixed interaction budget. BaRA also differs from general web-navigation agents by validating extracted multimodal artifacts through rule-based provenance and accessibility checks.

\paragraph{Runtime Platform and Positioning.}
At the system level, BaRA is implemented on top of Browser-use \citep{browseruse2024}. Browser-use is a general-purpose web-agent runtime that directly controls live browser sessions and can be paired with diverse LLM backends, enabling realistic evaluation on live websites. On top of this runtime, BaRA complements existing approaches as a collection-oriented control layer that orchestrates BFS-based link discovery under a fixed interaction budget, per-page extraction constraints, reflection, and download validation.

\section{BaRA Agent}
\label{sec:method}

BaRA is built around three design principles for budget-constrained, site-level multimodal web data collection: deterministic BFS-based link discovery for predictable site-internal coverage, liveness verification for filtering hallucinated and dead links, and history-based self-reflection with artifact verification for improving collection reliability.

\subsection{Problem Setup}
\label{sec:notation}

Given a seed URL $u_0$, BaRA crawls site-internal pages under a fixed interaction budget $(D,W,P,A)$, denoting the maximum BFS depth, retained child links per page, visited pages, and retry attempts, respectively. The visited URL sequence is $U=(u_0,\ldots,u_{m-1})$, where $m\leq P$. For each page $u_i$, BaRA extracts normalized and filtered internal links $C(u_i)$ and expands only a subset $\hat{C}(u_i)\subseteq C(u_i)$ with $|\hat{C}(u_i)|\leq W$. The objective is to return provenance-grounded text, image, and video artifacts from pages discovered within this budget.

\subsection{BFS-Based Link Discovery with Liveness Verification}
\label{sec:bfs}

Prompt-based web agents often rely on the LLM to decide which links to follow. This causes the agent to over-commit to a seemingly promising deep branch, leaving substantial parts of the website unexplored. BaRA addresses this issue with BFS. By expanding the frontier in depth order using a first-in, first-out (FIFO) queue, BFS prevents the agent from following a single branch too deeply and makes site-internal coverage predictable under the page budget $P$.

However, even with BFS, reliable link discovery requires addressing two additional failure modes. First, if link extraction is delegated to an LLM, the agent may generate URLs that do not exist, resulting in hallucinated links. Second, even when links are extracted from the Document Object Model (DOM), some anchors may point to pages that return HTTP errors such as 404 or 410. These dead links waste the fixed interaction budget by sending the crawler to unreachable pages. BaRA filters both types of failures as follows.

\paragraph{Filtering hallucinated links.}
BaRA removes the LLM from link extraction. For each page, it renders the page with Playwright \citep{microsoft_playwright} and directly extracts HTML anchor elements, i.e., \texttt{<a href>}, from the DOM to construct $C(u_i)$. Since candidate links are obtained from the rendered page rather than generated by the LLM, BaRA avoids hallucinated URLs during link discovery.

\paragraph{Filtering dead links.}
BaRA next applies two-stage liveness verification to candidate child links. It considers anchors in DOM order and keeps at most $W$ live child links for BFS expansion. BaRA first probes each candidate with an HTTP request and then verifies browser accessibility with Playwright before enqueueing it. If a candidate is filtered out, BaRA continues to the next anchor in DOM order until either $W$ live child links are retained or no candidates remain. Only live child links that satisfy the depth constraint $D$ and have not been visited are added to the BFS queue. The link discovery pass terminates when BaRA has visited $P$ pages or when the queue becomes empty.

\begin{figure*}[t]
    \centering
    \includegraphics[width=\textwidth]{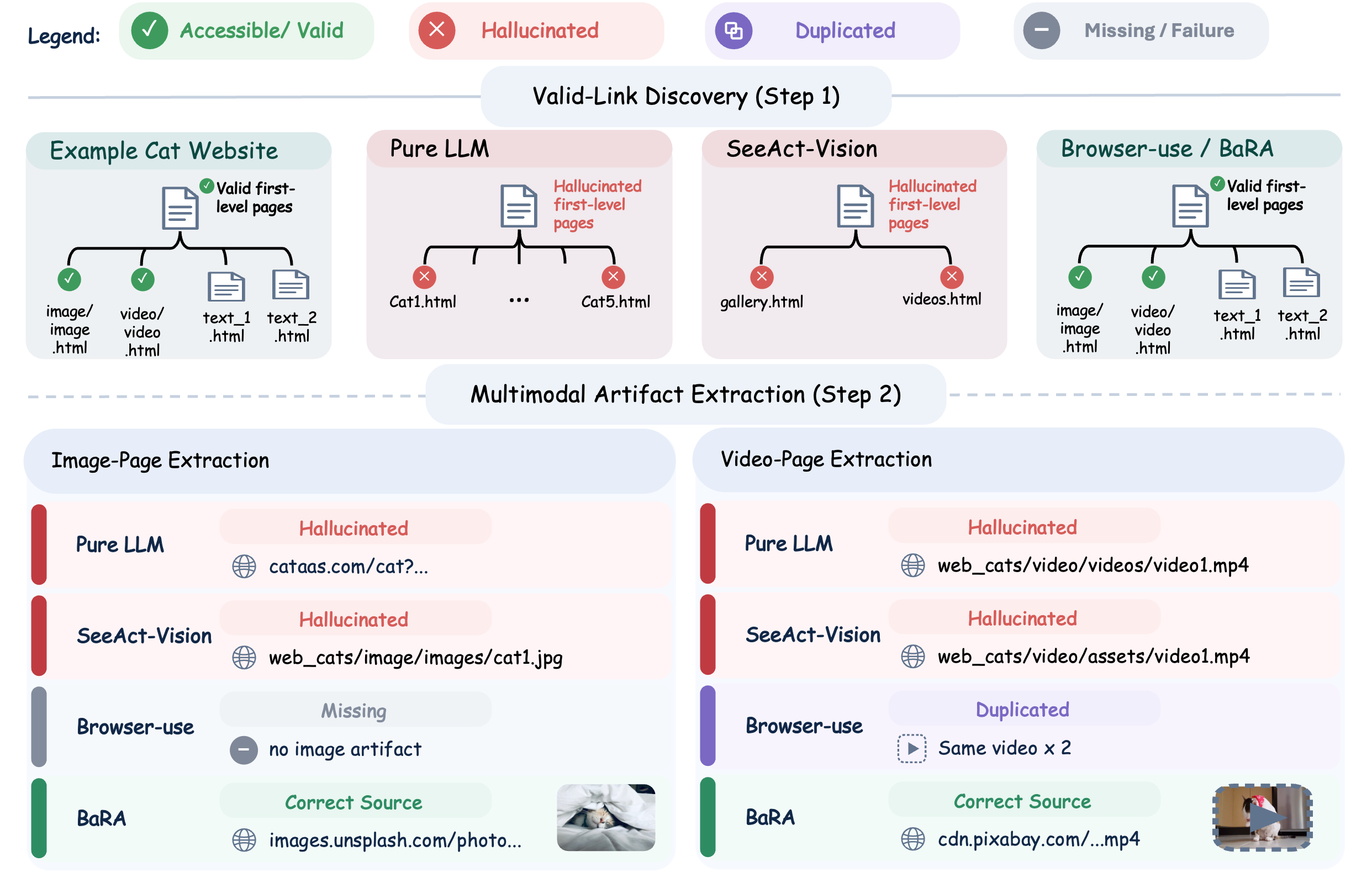}
    \caption{
Illustration of the synthetic benchmark and evaluation protocol.
\textbf{Top: valid-link discovery} evaluates whether discovered URLs pass two-stage liveness verification.
\textbf{Bottom: multimodal artifact extraction} evaluates whether extracted text, image, and video artifacts are page-grounded and accessible.
The example shows representative failures from baselines and the corresponding BaRA outputs.
}

    \label{fig:qualitative-case}
\end{figure*} 

\subsection{Reliable Multimodal Extraction via Self-Reflection and Artifact Verification}
\label{sec:reliability}

Live web data collection is vulnerable to unpredictable execution failures caused by dynamic DOM updates, pop-ups, rendering errors, and heterogeneous page layouts. Even when execution succeeds, the extracted artifacts are not necessarily reliable: a returned URL may not have appeared on the source page, may have the wrong format, may be non-downloadable, or may duplicate an artifact already collected from the same site. Text outputs may also contain hallucinated sentences or summaries that are not grounded in the page DOM. Thus, returning a raw list of URLs or text spans is insufficient for constructing a reusable dataset. BaRA addresses these issues with history-based self-reflection and rule-based artifact verification.

\paragraph{Recovering from execution failures with reflection.}
Agent-based extraction is brittle under dynamic page structures and unexpected rendering failures, and a single attempt may return incomplete outputs. During multimodal extraction, if the agent fails or returns an incomplete result, BaRA diagnoses the failure using the execution history $h_t$ and the partial output $y_t$, and revises the instruction for the next attempt:
\[
p_{t+1} \leftarrow R(p_t, h_t, y_t).
\]
where $t \in \{1, \ldots, A\}$ is the attempt index. The reflection module $R(\cdot)$ is not a blind retry mechanism. Instead, it updates the prompt according to the observed failure pattern, reducing the chance of repeating the same error. BaRA repeats this process for up to $A$ attempts and merges valid outputs across attempts.

\paragraph{Validating artifacts with provenance and accessibility checks.}
Successful extraction does not by itself guarantee artifact validity. For image and video artifacts, BaRA applies a sequence of rule-based checks before including them in the final dataset. It verifies whether the artifact was actually observed on the source page, whether the URL is externally accessible and downloadable, whether the downloaded content matches the expected modality and file format, and whether the artifact is duplicated within the same site. These checks ensure that image and video artifacts are not merely plausible URLs, but download-valid outputs with explicit provenance.

Text artifacts require a different verification procedure because they are textual spans extracted from the source page rather than downloadable objects. BaRA first normalizes the source-page DOM text and checks whether the extracted text appears in the normalized DOM. If substring matching fails, BaRA applies a token-level Jaccard similarity fallback to tolerate minor formatting differences while still rejecting unsupported or hallucinated text. 

For every artifact, BaRA records whether it passes each verification gate and, if excluded, the reason for exclusion. Only artifacts satisfying the modality-specific inclusion criteria are retained. As a result, BaRA outputs not a raw list of URLs or text snippets, but a traceable and verifiable multimodal dataset whose artifacts are provenance-grounded and available in an accessible form.

\section{Benchmarks}
\label{sec:benchmarks}

We evaluate BaRA on two complementary benchmarks for budget-constrained, site-level multimodal web data collection: a controlled synthetic benchmark with deterministic site structure and artifact targets, and a real-world benchmark constructed from live websites. The synthetic benchmark provides a repeatable evaluation setting unaffected by website updates, whereas the real-world benchmark evaluates valid-link discovery and download-valid multimodal extraction under heterogeneous live-website conditions.

\subsection{Synthetic Benchmark}
\label{sec:benchmark-synthetic}

To obtain a controlled environment, we construct a synthetic benchmark consisting of 50 topic-specific static websites. Unlike live websites, the synthetic benchmark has fixed link graphs and media assets, enabling repeatable evaluation without nondeterministic changes from website updates. The benchmark is designed to evaluate site-internal page discovery and accessible retrieval of text, image, and video artifacts under a fixed interaction budget.

\paragraph{Site construction.}
We first generate 50 topics with an LLM and fix them in a JSON specification. Each topic is instantiated as a static website using a shared family of HTML5 templates. All site-internal links are represented with relative paths, styles are local to each page, and no external JavaScript is included. The templates vary in structure, ranging from single-page sites to multi-level sites with portal-style entry pages. This design ensures that evaluation difficulty comes from site structure and content organization rather than client-side execution.

\paragraph{Content generation.}
Topic-specific text and captions are generated with Google Gemini \citep{googleaidev2026gemini3flashpreview}, while image and video assets are retrieved from the Unsplash and Pixabay APIs \citep{unsplash_api, pixabay_api} after URL validation. Thus, the text layer is generated with model assistance, whereas the media layer is grounded in explicitly verified external URLs. We allow small variations in the number of media assets and text subpages across topics to prevent reliance on a fixed cardinality pattern.

\paragraph{Deterministic structure and artifact targets.}
Because the synthetic corpus is static, its site structure and artifact targets are deterministically reconstructed from the generated site directories. For valid-link discovery, we evaluate the accessibility of discovered URLs using the same liveness-based protocol as in BaRA, rather than requiring an exact match to a reference crawl. For multimodal artifact extraction, target artifacts are derived from the generated page content and embedded valid image and video URLs. Authentication, form submission, and client-side rendering are intentionally excluded from the synthetic benchmark. Additional construction details are provided in Appendix~\ref{sec:appendix-select website}.

Figure~\ref{fig:qualitative-case} provides a running example of the synthetic benchmark. The top panel illustrates Step~1 valid-link discovery: agents may output semantically plausible URLs, but a URL is counted as valid only if it passes two-stage liveness verification. The bottom panel illustrates Step~2 multimodal artifact extraction, where predicted media artifacts are judged by whether they are grounded in the source page and available in an accessible form. We use this example only to clarify the evaluation protocol; aggregate results are reported in Section~\ref{sec:results}.

\subsection{Real-World Benchmark}
\label{sec:benchmark-realworld}

Existing evaluations of web crawling agents often rely on a small number of large domains or manually curated site lists, which risk underrepresenting the topical and structural diversity of the live web. To evaluate BaRA in a more heterogeneous setting, we construct a real-world benchmark consisting of 50 English public websites automatically collected and filtered from live web sources. The goal is to evaluate site-internal page discovery and provenance-grounded multimodal artifact retrieval under realistic website conditions.

\paragraph{Collection.}
We collect candidate websites from two independent sources: Tranco long-tail domains and Hacker News. For Tranco, we sample domains from ranks 200k--800k, avoiding both highly ranked domains that are often dominated by complex single-page applications and low-ranked domains that are more likely to be inactive. For Hacker News, we collect stories from the last 12 months with at least 20 points, using the community voting score as a proxy for attention.

\paragraph{Filtering.}
Each candidate website is filtered through a sequence of accessibility and content checks. We require that the site allow access by our identifying user-agent under \texttt{robots.txt}, return an HTML body, contain English content, and expose content without login forms or subscription walls. We require that at least one image or video be discoverable within two link hops from the entry URL. This final condition ensures that each website supports the multimodal extraction task.

\paragraph{Diversity constraints.}
To reduce topical and domain-type concentration, we apply a topic-balanced selection procedure. We build TF-IDF vectors from each candidate site's title and meta description, cluster candidates with KMeans using $K=12$, and select 50 websites while balancing TLD and topic-cluster distributions. The final benchmark contains 26 Tranco-originated sites and 24 Hacker-News-originated sites. It spans 19 TLDs, including \texttt{.com}, \texttt{.io}, \texttt{.dev}, \texttt{.org}, \texttt{.ca}, \texttt{.law}, and \texttt{.edu}. Each of the 12 topic clusters contains 4--5 websites, covering a range of live web topics such as technology, media, law, education, and commerce. The selected real-world websites are listed in Appendix~\ref{sec:appendix-realworld-selection}.

\section{Experimental Setup}
\label{sec:experimental-setup}

We evaluate BaRA as a reliable live web data collection pipeline under a fixed interaction budget. The evaluation focuses on two end-to-end capabilities: \emph{(i) valid-link discovery} (Step~1), which measures whether discovered URLs are actually accessible, and \emph{(ii) provenance-grounded artifact extraction} (Step~2), which measures whether extracted text, image, and video artifacts are valid and usable. Given a seed URL, each system first discovers site-internal URLs and then extracts raw text and image/video URLs from the visited pages in a predefined output format. Separate evaluation of the two steps distinguishes exploration failures from extraction failures. The single-page extraction prompt for Step~2 is provided in Appendix~\ref{app:single-page-extraction-prompt}.

\subsection{Budgets and Interaction Constraints}
\label{sec:budgets}

All methods share the same maximum page budget of $P=50$ visited pages and use \texttt{gemini-3-flash-preview} as the backbone model. Pure LLM, SeeAct, and Browser-use operate with generalized prompt-based exploration instructions. In contrast, BaRA performs DOM-based BFS link discovery without using the LLM to generate or choose links. We set BaRA's maximum BFS depth to $D=3$ and the maximum number of retained child links per page to $W=5$.

BaRA’s BFS-based link discovery is implemented as a deterministic DOM-based procedure and is not directly transferable to prompt-based baselines. We therefore do not force baselines to follow BaRA's link discovery strategy. Instead, we provide them with a generalized bounded-exploration prompt, which matches the standard use of prompt-based web agents. The prompt details are provided in Appendix~\ref{sec:appendix-generalized-prompts}. We set the maximum number of reflection-based retry attempts to $A=2$. Additional runtime settings, including timeouts, are reported in Appendix~\ref{app:runtime-configuration}.

\subsection{Baselines}
\label{sec:baselines}

We compare BaRA against three baselines.

\begin{itemize}
    \item \textbf{Pure LLM}: a baseline that uses only the backbone model to propose links and artifacts.

    \item \textbf{SeeAct-Vision} \citep{zheng2024gpt}: a vision-based web agent that observes webpage screenshots and grounds actions to HTML elements, representing an interactive agent for dynamic live websites.

    \item \textbf{Browser-use} \citep{browseruse2024}: a general-purpose browser agent runtime that supports different LLM backends and serves as a representative baseline for interaction-based live web data collection.
\end{itemize}

\begin{table}[t]
\caption{
Step~1 valid-link discovery results on the synthetic and real-world benchmarks.
Micro average pools discovered URLs across websites, while macro average averages site-level valid-link ratios.
}
\label{tab:step1}
\centering

\resizebox{\linewidth}{!}{
\begin{tabular}{@{}lcccc@{}}
\toprule
& \multicolumn{2}{c}{\textbf{Synthetic}} 
& \multicolumn{2}{c}{\textbf{Real-world}} \\
\cmidrule(lr){2-3}
\cmidrule(l){4-5}
\textbf{Method}
& \textbf{Micro avg.}
& \textbf{Macro avg.}
& \textbf{Micro avg.}
& \textbf{Macro avg.} \\
\midrule

Pure LLM    
& 0.1323 & 0.2263 
& 0.2543 & 0.2492 \\

SeeAct      
& 0.4583 & 0.5481 
& 0.9825 & 0.9208 \\

Browser-use 
& 0.9487 & 0.9656 
& 0.9396 & 0.9342 \\

\textbf{BaRA}
& \textbf{1.0000} & \textbf{1.0000}
& \textbf{0.9964} & \textbf{0.9388} \\

\bottomrule
\end{tabular}
}

\end{table}

\begin{table*}[t]
\caption{
Modality-wise Precision, Recall, and Accuracy under the \textbf{end-to-end setting}.
Results are grouped by dataset type, i.e., \textbf{synthetic} and \textbf{real-world}, and then by modality.
Prec., Rec., and Acc. denote Precision, Recall, and Accuracy, respectively.
}
\label{tab:step2_e2e}
\centering
\scriptsize
\setlength{\tabcolsep}{1.6pt}
\renewcommand{\arraystretch}{1.08}

\begin{tabular}{@{}l
ccc@{\hspace{3pt}}ccc@{\hspace{3pt}}ccc
@{\hspace{7pt}}
ccc@{\hspace{3pt}}ccc@{\hspace{3pt}}ccc
@{}}
\toprule
& \multicolumn{9}{c}{\textbf{Synthetic}}
& \multicolumn{9}{c}{\textbf{Real-world}} \\
\cmidrule(lr){2-10}
\cmidrule(lr){11-19}

& \multicolumn{3}{c}{\textbf{Text}}
& \multicolumn{3}{c}{\textbf{Image}}
& \multicolumn{3}{c}{\textbf{Video}}
& \multicolumn{3}{c}{\textbf{Text}}
& \multicolumn{3}{c}{\textbf{Image}}
& \multicolumn{3}{c}{\textbf{Video}} \\
\cmidrule(lr){2-4}
\cmidrule(lr){5-7}
\cmidrule(lr){8-10}
\cmidrule(lr){11-13}
\cmidrule(lr){14-16}
\cmidrule(lr){17-19}

\textbf{Method}
& \textbf{Prec.} & \textbf{Rec.} & \textbf{Acc.}
& \textbf{Prec.} & \textbf{Rec.} & \textbf{Acc.}
& \textbf{Prec.} & \textbf{Rec.} & \textbf{Acc.}
& \textbf{Prec.} & \textbf{Rec.} & \textbf{Acc.}
& \textbf{Prec.} & \textbf{Rec.} & \textbf{Acc.}
& \textbf{Prec.} & \textbf{Rec.} & \textbf{Acc.} \\
\midrule

Pure LLM
& 0.0965 & 0.1149 & 0.0454
& 0.0032 & 0.0028 & 0.0017
& 0.0000 & 0.0000 & 0.0000
& 0.1579 & 0.2356 & 0.0749
& 0.0163 & 0.0175 & 0.0154
& 0.0000 & 0.0000 & 0.0000 \\

SeeAct
& 0.4206 & 0.3283 & 0.3070
& 0.0000 & 0.0000 & 0.0000
& 0.0000 & 0.0000 & 0.0000
& 0.8211 & 0.4580 & 0.4330
& 0.0242 & 0.0178 & 0.0142
& 0.0000 & 0.0000 & 0.0000 \\

Browser-use
& 0.9592 & 0.9406 & 0.9350
& 0.9184 & 0.9184 & 0.9184
& 0.5162 & 0.6429 & 0.5162
& \textbf{0.9181} & 0.7442 & 0.6982
& 0.8072 & 0.6013 & 0.5829
& 0.0140 & 0.0242 & 0.0111 \\

\textbf{BaRA}
& \textbf{0.9726} & \textbf{0.9790} & \textbf{0.9522}
& \textbf{1.0000} & \textbf{0.9564} & \textbf{0.9564}
& \textbf{1.0000} & \textbf{0.9796} & \textbf{0.9796}
& 0.9172 & \textbf{0.7513} & \textbf{0.7061}
& \textbf{0.9177} & \textbf{0.7728} & \textbf{0.7639}
& \textbf{0.2468} & \textbf{0.2321} & \textbf{0.2321} \\

\bottomrule
\end{tabular}

\end{table*}

\begin{table}[t]
\caption{
Modality-wise Precision, Recall, and Accuracy under the \textbf{shared-page} setting on real-world pages commonly reached by all methods.}
\label{tab:step2_shared}
\centering
\scriptsize
\setlength{\tabcolsep}{1.6pt}
\renewcommand{\arraystretch}{1.05}

\resizebox{\columnwidth}{!}{%
\begin{tabular}{@{}lccc@{\hspace{3pt}}ccc@{\hspace{3pt}}ccc@{}}
\toprule
& \multicolumn{3}{c}{\textbf{Text}}
& \multicolumn{3}{c}{\textbf{Image}}
& \multicolumn{3}{c}{\textbf{Video}} \\
\cmidrule(lr){2-4}
\cmidrule(lr){5-7}
\cmidrule(lr){8-10}

\textbf{Method}
& \textbf{Prec.} & \textbf{Rec.} & \textbf{Acc.}
& \textbf{Prec.} & \textbf{Rec.} & \textbf{Acc.}
& \textbf{Prec.} & \textbf{Rec.} & \textbf{Acc.} \\
\midrule

Pure LLM
& 0.4747 & 0.1730 & 0.1375
& 0.0038 & 0.0038 & 0.0023
& 0.0000 & 0.0000 & 0.0000 \\

SeeAct
& 0.8905 & 0.4973 & 0.4646
& 0.0030 & 0.0009 & 0.0007
& 0.0000 & 0.0000 & 0.0000 \\

Browser-use
& 0.6092 & 0.4935 & 0.4811
& 0.5145 & 0.3422 & 0.3336
& 0.0882 & 0.1176 & 0.0882 \\

\textbf{BaRA}
& \textbf{0.9179} & \textbf{0.7388} & \textbf{0.7095}
& \textbf{0.7019} & \textbf{0.4712} & \textbf{0.4689}
& \textbf{0.7500} & \textbf{0.7500} & \textbf{0.7500} \\

\bottomrule
\end{tabular}%
}

\end{table}
\subsection{Evaluation Metrics}
\label{sec:metrics}

For Step~1, we evaluate valid-link discovery. For each URL discovered by an agent, we perform two-stage liveness verification using an HTTP request followed by a Playwright-based accessibility check. A discovered URL is counted as valid only if it passes both checks. We report both micro and macro averages. The micro average computes the proportion of valid URLs after pooling discovered URLs across all websites, while the macro average first computes the valid-link ratio for each website and then averages the ratios across websites.

For Step~2, we compare predicted and reference artifact sets. Let $S_p$ and $S_r$ be the predicted and reference sets, and let $I=|S_p\cap S_r|$. We compute
\[
\begin{array}{c}
\mathrm{Prec}=\frac{I}{|S_p|}, \quad
\mathrm{Rec}=\frac{I}{|S_r|}, \quad
\mathrm{Acc}=\frac{I}{|S_p\cup S_r|}.
\end{array}
\]
Prec, Rec, and Acc denote Precision, Recall, and Accuracy, respectively; Accuracy is the Jaccard index. Metrics are computed separately for text, images, and videos, using normalized word sets for text and normalized URL sets for images and videos. Details are provided in Appendix~\ref{sec:app-metrics}; unless otherwise stated, we report end-to-end results.

\section{Results and Analysis}
\label{sec:results}

\subsection{Main Results}
\label{sec:main-results}

\paragraph{Valid-link discovery.}
Table~\ref{tab:step1} reports Step~1 valid-link discovery results. BaRA achieves the highest valid-link ratio on both the synthetic and real-world benchmarks. On the synthetic benchmark, BaRA obtains perfect averages of 1.0000, indicating that all discovered URLs pass the two-stage liveness verification. On the real-world benchmark, BaRA also remains highly reliable. Browser-use is the strongest baseline on the synthetic benchmark and by macro average on the real-world benchmark, but still falls behind BaRA. SeeAct is relatively stable on real websites but drops substantially on the synthetic benchmark, while Pure LLM performs worst in both settings. These results show that browser-grounded interaction helps, but that DOM-based BFS link discovery with liveness verification is crucial for avoiding hallucinated and dead links under a fixed budget.

\paragraph{Multimodal artifact extraction.}
Table~\ref{tab:step2_e2e} reports Step~2 extraction results for text, images, and videos. BaRA achieves the best or near-best performance across modalities on both benchmarks. On the synthetic benchmark, BaRA reaches near-perfect image and video extraction, whereas Pure LLM and SeeAct largely fail to recover valid media artifacts. On the real-world benchmark, text extraction performance is closer between BaRA and Browser-use, but BaRA shows clearer gains for images and especially videos. This suggests that rule-based artifact verification is important for live web media, where extracted URLs may be inaccessible, duplicated, or not directly downloadable. Since the number of text spans and media artifacts varies across websites, Step~2 reports macro averages to avoid over-weighting a small number of large sites.
\begin{table*}[t]
\caption{
Ablation on the synthetic benchmark: valid-link discovery and multimodal extraction.
}
\label{tab:ablation-synthetic}
\centering
\scriptsize
\renewcommand{\arraystretch}{1.08}

\begin{subtable}[t]{0.29\textwidth}

\centering
\setlength{\tabcolsep}{4.0pt}
\begin{tabular}{@{}lcc@{}}
\toprule
\textbf{Variant} & \textbf{Micro avg.} & \textbf{Macro avg.}\\
\midrule

w/o BFS
& 0.7016 & 0.8074 \\

w/o liveness verification
& 0.9799 & 0.9810 \\

\textbf{Full BaRA}
& \textbf{1.0000} & \textbf{1.0000} \\

\bottomrule
\end{tabular}
\caption{Step-1 valid-link discovery.}
\label{tab:ablation-link-synthetic}
\end{subtable}
\hfill
\begin{subtable}[t]{0.69\textwidth}
\centering
\setlength{\tabcolsep}{2.0pt}
\begin{tabular}{@{}lccccccccc@{}}
\toprule
& \multicolumn{3}{c}{\textbf{Text}}
& \multicolumn{3}{c}{\textbf{Image}}
& \multicolumn{3}{c}{\textbf{Video}} \\
\cmidrule(lr){2-4}
\cmidrule(lr){5-7}
\cmidrule(l){8-10}
\textbf{Variant}
& \textbf{Prec.} & \textbf{Rec.} & \textbf{Acc.}
& \textbf{Prec.} & \textbf{Rec.} & \textbf{Acc.}
& \textbf{Prec.} & \textbf{Rec.} & \textbf{Acc.}\\
\midrule

w/o reflection
& 0.6099 & 0.6094 & 0.5923
& 1.0000 & 0.9496 & 0.9496
& 0.9783 & 0.9565 & 0.9565 \\

w/o retry-and-merge
& 0.9335 & 0.9362 & 0.9112
& 0.9796 & 0.9356 & 0.9356
& 1.0000 & 0.9787 & 0.9787 \\

w/o artifact verification
& 0.9694 & 0.9768 & 0.9470
& 0.8378 & 0.8367 & 0.8367
& 0.6565 & 0.6463 & 0.6463 \\

\textbf{Full BaRA}
& \textbf{0.9726} & \textbf{0.9790} & \textbf{0.9522}
& \textbf{1.0000} & \textbf{0.9564} & \textbf{0.9564}
& \textbf{1.0000} & \textbf{0.9796} & \textbf{0.9796} \\

\bottomrule
\end{tabular}
\caption{Step-2 multimodal extraction.}
\label{tab:ablation-extraction-synthetic}
\end{subtable}

\end{table*}

\begin{table*}[t]
\caption{
GPT-4.1-mini results on the synthetic benchmark: valid-link discovery and multimodal extraction.
}
\label{tab:gpt41mini-synthetic-results}
\centering
\scriptsize
\renewcommand{\arraystretch}{1.08}

\begin{subtable}[t]{0.30\textwidth}
\centering
\setlength{\tabcolsep}{4.0pt}
\begin{tabular}{@{}lcc@{}}
\toprule
\textbf{Method} & \textbf{Micro avg.} & \textbf{Macro avg.} \\
\midrule

Pure LLM
& 0.1149 & 0.2005 \\

SeeAct
& 0.9951 & 0.9950 \\

Browser-use
& 0.4564 & 0.4728 \\

\textbf{BaRA}
& \textbf{1.0000} & \textbf{1.0000} \\

\bottomrule
\end{tabular}
\caption{Step-1 valid-link discovery.}
\label{tab:gpt41mini-step1}
\end{subtable}
\hfill
\begin{subtable}[t]{0.68\textwidth}
\centering
\setlength{\tabcolsep}{2.0pt}
\resizebox{\linewidth}{!}{
\begin{tabular}{@{}lccccccccc@{}}
\toprule
& \multicolumn{3}{c}{\textbf{Text}}
& \multicolumn{3}{c}{\textbf{Image}}
& \multicolumn{3}{c}{\textbf{Video}} \\
\cmidrule(lr){2-4}
\cmidrule(lr){5-7}
\cmidrule(l){8-10}
\textbf{Method}
& \textbf{Prec.} & \textbf{Rec.} & \textbf{Acc.}
& \textbf{Prec.} & \textbf{Rec.} & \textbf{Acc.}
& \textbf{Prec.} & \textbf{Rec.} & \textbf{Acc.}\\
\midrule

Pure LLM
& 0.2305 & 0.1591 & 0.0925
& 0.0000 & 0.0000 & 0.0000
& 0.0000 & 0.0000 & 0.0000 \\

SeeAct
& 0.8722 & 0.5969 & 0.5779
& 0.0000 & 0.0000 & 0.0000
& 0.0000 & 0.0000 & 0.0000 \\

Browser-use
& 0.4761 & 0.4661 & 0.4586
& 0.1786 & 0.1786 & 0.1786
& 0.1871 & 0.1799 & 0.1799 \\

\textbf{BaRA}
& \textbf{0.9729} & \textbf{0.9829} & \textbf{0.9564}
& \textbf{0.9796} & \textbf{0.9770} & \textbf{0.9770}
& \textbf{0.9796} & \textbf{0.9592} & \textbf{0.9592} \\

\bottomrule
\end{tabular}
}
\caption{Step-2 multimodal extraction.}
\label{tab:gpt41mini-step2}
\end{subtable}

\end{table*}

\begin{table*}[!t]
\caption{
Modality-wise precision, recall, and accuracy on the synthetic benchmark under the recovery setting.
Recovery denotes the use of reflection and retry-and-merge.
Best scores in each column are shown in bold.
}
\label{tab:synthetic-recovery}
\centering
\scriptsize
\renewcommand{\arraystretch}{1.08}
\setlength{\tabcolsep}{3.0pt}
\begin{tabular}{@{}lccccccccc@{}}
\toprule
& \multicolumn{3}{c}{\textbf{Text}}
& \multicolumn{3}{c}{\textbf{Image}}
& \multicolumn{3}{c}{\textbf{Video}} \\
\cmidrule(lr){2-4}
\cmidrule(lr){5-7}
\cmidrule(l){8-10}
\textbf{Variant}
& \textbf{Prec.} & \textbf{Rec.} & \textbf{Acc.}
& \textbf{Prec.} & \textbf{Rec.} & \textbf{Acc.}
& \textbf{Prec.} & \textbf{Rec.} & \textbf{Acc.} \\
\midrule

Pure LLM + Recovery
& 0.1031 & 0.1035 & 0.0460
& 0.0000 & 0.0000 & 0.0000
& 0.0000 & 0.0000 & 0.0000 \\

SeeAct + Recovery
& 0.8546 & 0.8446 & 0.8281
& 0.4567 & 0.4567 & 0.4567
& 0.4567 & 0.4567 & 0.4567 \\

Browser-use + Recovery
& \textbf{0.9945} & 0.9761 & \textbf{0.9707}
& 0.9324 & 0.9297 & 0.9297
& 0.6642 & 0.7503 & 0.6656 \\

BaRA
& 0.9726 & \textbf{0.9790} & 0.9522
& \textbf{1.0000} & \textbf{0.9564} & \textbf{0.9564}
& \textbf{1.0000} & \textbf{0.9796} & \textbf{0.9796} \\

\bottomrule
\end{tabular}

\end{table*}
\paragraph{Fair comparison of Step~2 extraction.}
End-to-end extraction results reflect both Step~1 link discovery and Step~2 multimodal artifact extraction. To isolate extraction quality, we additionally evaluate all methods on the set of pages reached by all four systems, as shown in Table~\ref{tab:step2_shared}. This shared-page setting removes differences caused by link discovery and compares the extraction component on the same page set. On the synthetic benchmark, the common page set does not contain image or video targets, making this analysis uninformative for multimodal extraction. We therefore report the shared-page analysis only on the real-world benchmark. Even on the same pages, BaRA achieves the best performance across all modalities, with the largest margin on video extraction.

\subsection{Ablation Analysis}
\label{sec:ablation}

We ablate BaRA on the synthetic benchmark to isolate the effects of its key design choices.

\paragraph{Step~1: BFS-based link discovery and liveness verification.} Table~\ref{tab:ablation-synthetic} shows that replacing BaRA's BFS-based link discovery with a generalized bounded-exploration prompt (\textit{w/o BFS}) substantially reduces the valid-link ratio from 1.000 to 0.7016.
Since BaRA's BFS-based link discovery is coupled with deterministic DOM-based link extraction, this ablation removes both systematic breadth-first exploration and hallucination-free link extraction. Disabling liveness verification (\textit{w/o liveness verification}) causes only a small drop on the synthetic benchmark, because the static synthetic sites contain few inaccessible links by construction. This limited drop reflects the synthetic benchmark construction, where inaccessible links are rare; live websites contain more dead-link cases.

\paragraph{Step~2: reflection and artifact verification.}
The Step~2 ablations show that BaRA's reliability components contribute differently across modalities. Removing reflection (\textit{w/o reflection}) substantially reduces text accuracy from 0.9522 to 0.5923, while image and video extraction are much less affected. In contrast, removing artifact verification (\textit{w/o artifact verification}) has a limited effect on text but reduces video accuracy from 0.9796 to 0.6463. Disabling retry-and-merge (\textit{w/o retry-and-merge}) slightly lowers text performance and has little effect on image and video extraction. These results suggest a complementary division of labor: reflection mainly improves the stability of text extraction, whereas artifact verification ensures media validity.

\paragraph{Backbone robustness.}
We test backbone robustness on the synthetic benchmark using GPT-4.1-mini \citep{openai2025gpt41mini} (Table~\ref{tab:gpt41mini-synthetic-results}). Because Step~1 does not use the LLM to generate or choose links, BaRA remains at 1.0000 across backbones, whereas LLM-based baselines vary more. In Step~2, BaRA remains strong across text, image, and video, and the pattern remains: BaRA is stable across backbones, while LLM-dependent baselines vary more. These results indicate that BaRA's gains stem primarily from its collection-oriented control design rather than from a particular backbone.


\paragraph{Applying recovery to baselines.}
Table~\ref{tab:synthetic-recovery} reports the effect of applying recovery to baseline agents outside BaRA. Here, recovery denotes reflection with retry-and-merge. Recovery improves SeeAct and Browser-use in this setting, indicating that history-based retry contributes to robustness across the evaluated agents. The gain is particularly large for SeeAct: its text accuracy increases from 0.3070 in the main end-to-end results (Table~\ref{tab:step2_e2e}) to 0.8281 with recovery. In contrast, Pure LLM remains low even after adding recovery, suggesting that recovery alone is insufficient without browser-grounded interaction and page-level evidence.

Despite these gains, the recovered baselines still fall short of BaRA, especially for image and video extraction. This suggests that BaRA's advantage does not come from the recovery module alone. Rather, it comes from the joint design of BFS-based link discovery, liveness verification, reflection, and rule-based artifact verification for budget-constrained, site-level multimodal web data collection.

\section{Conclusion}

We present BaRA, a framework for budget-constrained, multimodal web data collection. BaRA casts collection under a fixed budget as verified exploration, using BFS-based link discovery, two-stage liveness verification, reflection, and artifact verification to obtain provenance-grounded multimodal artifacts. We introduce synthetic and real-world benchmarks and show that BaRA consistently improves valid-link discovery and download-valid multimodal extraction over existing web agents. The results suggest that reliable live web data collection requires evaluating not only task completion, but also whether limited interactions yield accessible and verifiable multimodal artifacts.

BaRA is limited to publicly accessible content under bounded interactions and may miss content behind hidden UI states, authentication, or JavaScript-heavy workflows. Deterministic BFS may also omit links beyond its depth and width budgets, motivating adaptive prioritization. Our benchmarks exclude paywalled and highly interactive sites. Deployment should respect site policies, with outputs reviewed for copyright, privacy, and downstream use.

\section*{Acknowledgements}

This work was supported by the Starting Growth Technological R\&D Program (TIPS Program, (No.~RS-2024-0050914)) funded by the Ministry of SMEs and Startups (MSS, Korea) in 2024. This work was supported by the Institute of Information \& Communications Technology Planning \& Evaluation (IITP) grant funded by the Korean government (MSIT) (No.~RS-2023-00224740, Development of technology to prevent and track the distribution of illegally filmed content).

\bibliography{custom}

@article{nakano2021webgpt,
  title={Webgpt: Browser-assisted question-answering with human feedback},
  author={Nakano, Reiichiro and Hilton, Jacob and Balaji, Suchir and Wu, Jeff and Ouyang, Long and Kim, Christina and Hesse, Christopher and Jain, Shantanu and Kosaraju, Vineet and Saunders, William and others},
  journal={arXiv preprint arXiv:2112.09332},
  year={2021}
}

@article{deng2023mind2web,
  title={Mind2web: Towards a generalist agent for the web},
  author={Deng, Xiang and Gu, Yu and Zheng, Boyuan and Chen, Shijie and Stevens, Sam and Wang, Boshi and Sun, Huan and Su, Yu},
  journal={Advances in Neural Information Processing Systems},
  volume={36},
  pages={28091--28114},
  year={2023}
}

@article{zhou2023webarena,
  title={Webarena: A realistic web environment for building autonomous agents},
  author={Zhou, Shuyan and Xu, Frank F and Zhu, Hao and Zhou, Xuhui and Lo, Robert and Sridhar, Abishek and Cheng, Xianyi and Ou, Tianyue and Bisk, Yonatan and Fried, Daniel and others},
  journal={arXiv preprint arXiv:2307.13854},
  year={2023}
}

@inproceedings{he2024webvoyager,
  title={Webvoyager: Building an end-to-end web agent with large multimodal models},
  author={He, Hongliang and Yao, Wenlin and Ma, Kaixin and Yu, Wenhao and Dai, Yong and Zhang, Hongming and Lan, Zhenzhong and Yu, Dong},
  booktitle={Proceedings of the 62nd Annual Meeting of the Association for Computational Linguistics (Volume 1: Long Papers)},
  pages={6864--6890},
  year={2024}
}

@article{zheng2024gpt,
  title={Gpt-4v (ision) is a generalist web agent, if grounded},
  author={Zheng, Boyuan and Gou, Boyu and Kil, Jihyung and Sun, Huan and Su, Yu},
  journal={arXiv preprint arXiv:2401.01614},
  year={2024}
}

@inproceedings{yao2022react,
  title={React: Synergizing reasoning and acting in language models},
  author={Yao, Shunyu and Zhao, Jeffrey and Yu, Dian and Du, Nan and Shafran, Izhak and Narasimhan, Karthik R and Cao, Yuan},
  booktitle={The eleventh international conference on learning representations},
  year={2022}
}

@article{shinn2023reflexion,
  title={Reflexion: Language agents with verbal reinforcement learning},
  author={Shinn, Noah and Cassano, Federico and Gopinath, Ashwin and Narasimhan, Karthik and Yao, Shunyu},
  journal={Advances in neural information processing systems},
  volume={36},
  pages={8634--8652},
  year={2023}
}

@article{xue2025illusion,
  title={An illusion of progress? assessing the current state of web agents},
  author={Xue, Tianci and Qi, Weijian and Shi, Tianneng and Song, Chan Hee and Gou, Boyu and Song, Dawn and Sun, Huan and Su, Yu},
  journal={arXiv preprint arXiv:2504.01382},
  year={2025}
}

@misc{browseruse2024,
  author       = {{Browser-use contributors}},
  title        = {Browser-use: Open-source browser agent runtime for LLM-based web interaction},
  year         = {n.d.},
  howpublished = {\url{https://github.com/browser-use/browser-use}},
  note         = {GitHub repository. Accessed: 2026-05-26}
}

@inproceedings{shi2017world,
  title={World of bits: An open-domain platform for web-based agents},
  author={Shi, Tianlin and Karpathy, Andrej and Fan, Linxi and Hernandez, Jonathan and Liang, Percy},
  booktitle={International Conference on Machine Learning},
  pages={3135--3144},
  year={2017},
  organization={PMLR}
}

@article{liu2018reinforcement,
  title={Reinforcement learning on web interfaces using workflow-guided exploration},
  author={Liu, Evan Zheran and Guu, Kelvin and Pasupat, Panupong and Shi, Tianlin and Liang, Percy},
  journal={arXiv preprint arXiv:1802.08802},
  year={2018}
}

@article{yao2022webshop,
  title={Webshop: Towards scalable real-world web interaction with grounded language agents},
  author={Yao, Shunyu and Chen, Howard and Yang, John and Narasimhan, Karthik},
  journal={Advances in Neural Information Processing Systems},
  volume={35},
  pages={20744--20757},
  year={2022}
}

@inproceedings{koh2024visualwebarena,
  title={Visualwebarena: Evaluating multimodal agents on realistic visual web tasks},
  author={Koh, Jing Yu and Lo, Robert and Jang, Lawrence and Duvvur, Vikram and Lim, Ming and Huang, Po-Yu and Neubig, Graham and Zhou, Shuyan and Salakhutdinov, Russ and Fried, Daniel},
  booktitle={Proceedings of the 62nd Annual Meeting of the Association for Computational Linguistics (Volume 1: Long Papers)},
  pages={881--905},
  year={2024}
}

@article{drouin2024workarena,
  title={Workarena: How capable are web agents at solving common knowledge work tasks?},
  author={Drouin, Alexandre and Gasse, Maxime and Caccia, Massimo and Laradji, Issam H and Del Verme, Manuel and Marty, Tom and Boisvert, L{\'e}o and Thakkar, Megh and Cappart, Quentin and Vazquez, David and others},
  journal={arXiv preprint arXiv:2403.07718},
  year={2024}
}

@article{boisvert2024workarena++,
  title={Workarena++: Towards compositional planning and reasoning-based common knowledge work tasks},
  author={Boisvert, L{\'e}o and Thakkar, Megh and Gasse, Maxime and Caccia, Massimo and De Chezelles, Thibault L and Cappart, Quentin and Chapados, Nicolas and Lacoste, Alexandre and Drouin, Alexandre},
  journal={Advances in Neural Information Processing Systems},
  volume={37},
  pages={5996--6051},
  year={2024}
}

@article{lu2024weblinx,
  title={Weblinx: Real-world website navigation with multi-turn dialogue},
  author={L{\`u}, Xing Han and Kasner, Zden{\v{e}}k and Reddy, Siva},
  journal={arXiv preprint arXiv:2402.05930},
  year={2024}
}

@article{furuta2023multimodal,
  title={Multimodal web navigation with instruction-finetuned foundation models},
  author={Furuta, Hiroki and Lee, Kuang-Huei and Nachum, Ofir and Matsuo, Yutaka and Faust, Aleksandra and Gu, Shixiang Shane and Gur, Izzeddin},
  journal={arXiv preprint arXiv:2305.11854},
  year={2023}
}

@article{zheng2023synapse,
  title={Synapse: Trajectory-as-exemplar prompting with memory for computer control},
  author={Zheng, Longtao and Wang, Rundong and Wang, Xinrun and An, Bo},
  journal={arXiv preprint arXiv:2306.07863},
  year={2023}
}

@article{yang2024agentoccam,
  title={Agentoccam: A simple yet strong baseline for llm-based web agents},
  author={Yang, Ke and Liu, Yao and Chaudhary, Sapana and Fakoor, Rasool and Chaudhari, Pratik and Karypis, George and Rangwala, Huzefa},
  journal={arXiv preprint arXiv:2410.13825},
  year={2024}
}

@misc{firecrawl_repo,
  author       = {{Firecrawl}},
  title        = {{Firecrawl}},
  year         = {n.d.},
  howpublished = {\url{https://github.com/firecrawl/firecrawl}},
  note         = {GitHub repository. Accessed: 2026-05-26}
}

@misc{unsplash_api,
  author       = {Unsplash},
  title        = {Unsplash API Documentation},
  year         = {n.d.},
  howpublished = {\url{https://unsplash.com/documentation}},
  note         = {Accessed: 2026-05-26}
}

@misc{pixabay_api,
  author       = {Pixabay},
  title        = {Pixabay Developer API},
  year         = {n.d.},
  howpublished = {\url{https://pixabay.com/service/about/api/}},
  note         = {Accessed: 2026-05-26}
}

@manual{googleaidev2026gemini3flashpreview,
  key          = {Google AI for Developers},
  title        = {Gemini 3 Flash Preview},
  organization = {Google AI for Developers},
  year         = {2026},
  url          = {https://ai.google.dev/gemini-api/docs/models/gemini-3-flash-preview},
  note         = {Model code: \texttt{gemini-3-flash-preview}. Last updated 2026-02-19 UTC. Accessed: 2026-05-26.}
}

@misc{openai2025gpt41mini,
  author       = {{OpenAI}},
  title        = {{GPT-4.1 mini} Model},
  year         = {2025},
  howpublished = {OpenAI API documentation},
  url          = {https://developers.openai.com/api/docs/models/gpt-4.1-mini},
  note         = {Model code: \texttt{gpt-4.1-mini}; snapshot: \texttt{gpt-4.1-mini-2025-04-14}. Accessed: 2026-05-26}
}

@misc{microsoft_playwright,
  author       = {{Microsoft}},
  title        = {{Playwright}},
  howpublished = {\url{https://github.com/microsoft/playwright}},
  note         = {GitHub repository. Accessed: 2026-05-26},
  year         = {n.d.}
}

@article{pochat2018tranco,
  title={Tranco: A research-oriented top sites ranking hardened against manipulation},
  author={Pochat, Victor Le and Van Goethem, Tom and Tajalizadehkhoob, Samaneh and Joosen, Wouter and others},
  journal={arXiv preprint arXiv:1806.01156},
  year={2018}
}

@misc{algolia_hn_api,
  author       = {{Algolia}},
  title        = {{HN Search API}},
  howpublished = {\url{https://hn.algolia.com/api}},
  note         = {Accessed: 2026-05-26},
  year         = {n.d.}
}

@misc{openrouter_api,
  author       = {{OpenRouter}},
  title        = {{OpenRouter API Reference}},
  howpublished = {\url{https://openrouter.ai/docs/api/reference/overview}},
  note         = {Accessed: 2026-05-26},
  year         = {n.d.}
}
\bibliographystyle{icml2026}

\newpage
\appendix
\onecolumn
\section{Evaluation Details}
\label{sec:app-evaluation}

\subsection{Metric Details}
\label{sec:app-metrics}

\paragraph{Set-based metrics.}
For each modality (text, image, video) on each evaluated page, we compute Precision, Recall, and Accuracy, where Accuracy is the Jaccard index from the counts where $S_p$ is the predicted item set and $S_r$ is the reference set. For image and video, a page is excluded from the site-level average when both $S_p$ and $S_r$ are empty for that modality (i.e., the page contains no media of that type and none was predicted).
For text, by contrast, every page in the evaluation set is included in the site-level average. Any legitimate page should contain some visible text content, so an error page indicates that the corresponding site-internal URL was incorrectly extracted in Step~1 and itself constitutes a failure of the pipeline. Such pages contribute zero precision, zero recall, and zero Jaccard accuracy to the average.

\paragraph{Two-level macro aggregation.}
Step~2 extraction results are reported as two-level macro averages. Within each site, page-level metrics are averaged (macro) over the pages that have signal for the modality, yielding one site-level score per modality.  Those site-level scores are then averaged (macro) over all evaluated sites to produce the final reported number. This scheme treats every site equally regardless of how many pages it contributes and is consistent across the shared-page and end-to-end settings.

\paragraph{URL normalization for evaluation.}
Before set comparison, all URLs are mapped to a canonical form:
(i) scheme and host are lowercased;
(ii) the URL fragment is dropped;
(iii) trailing slashes and terminal \texttt{/index.\{html,htm,php\}} are stripped;
(iv) \emph{the entire query string is discarded}.
Step~(iv) is necessary because CDN-signed image URLs (e.g.\ Unsplash's \texttt{ixid} parameter) differ on every request, so retaining the query string would create systematic false negatives that do not reflect actual retrieval failure.

\paragraph{Text normalization for evaluation.}
For the text modality, reference and predicted strings are normalized before comparison: all characters are lowercased, punctuation is replaced by spaces, and whitespace runs are collapsed.  Text units are individual normalized \emph{words}, so a text snippet is modeled as a bag of words.

\subsection{Step~1: Link Normalization and Filtering}
\label{sec:app-step1}

The BFS frontier maintained by Step~1 applies the following pipeline to every anchor href extracted from a rendered page before the URL is considered for the queue.

\paragraph{File-extension and boilerplate-path exclusions.}
URLs whose path ends with a known binary or media extension are dropped unconditionally: archives (\texttt{.pdf}, \texttt{.zip}, \texttt{.rar}, \texttt{.7z}, \texttt{.apk}, \texttt{.dmg}, \texttt{.exe}), images (\texttt{.jpg}, \texttt{.jpeg}, \texttt{.png}, \texttt{.gif}, \texttt{.webp}, \texttt{.svg}), and video/streaming files (\texttt{.mp4}, \texttt{.mov}, \texttt{.avi}, \texttt{.wmv}, \texttt{.flv}, \texttt{.mkv}, \texttt{.webm}, \texttt{.m4v}, \texttt{.m3u8}, \texttt{.ts}).
Paths containing boilerplate segments are also excluded: \texttt{/privacy}, \texttt{/terms}, \texttt{/login}, \texttt{/signup}.

\paragraph{Liveness verification for dead-link filtering.}
After normalization and rule-based filtering, each candidate URL is first probed with an HTTP GET request. Candidates with dead-link responses are recorded but not enqueued. HTTP-live candidates are then rendered with Playwright to verify browser accessibility. The BFS width budget counts only links that pass both stages. When a live URL has undergone an HTTP redirect, the resolved (post-redirect) canonical URL is stored and enqueued instead of the original anchor.

\subsection{Step~2: Artifact Verification Pipeline}
\label{sec:app-verification}

In Step~2, BaRA applies a deterministic, LLM-free verification pipeline to the multimodal artifact candidates collected by the LLM-guided agent before retaining them into the final dataset. The verification procedure is defined according to the structural form of each modality. Images and videos are externally accessible URL objects, and are therefore verified through rule-based provenance, accessibility/downloadability, modality/file-format consistency, and site-level non-duplication checks. In contrast, text artifacts are not downloadable URLs but character strings extracted from the source page, and are therefore verified by checking whether they are grounded in the visible text of the source DOM. Accordingly, we define URL-level gates for image/video artifacts and separate text grounding checks for text artifacts.

\subsubsection{Image and Video Artifact Verification}

Image and video artifacts are represented as URLs and are verified through a five-stage URL-level pipeline. Under the default strict mode, an image or video candidate is included in the final dataset only if it passes all gates.

\paragraph{Gate~1 — Source-page provenance.}
After LLM-based extraction, BaRA re-renders the source page using a separate headless browser session and collects the rendered DOM and browser network log required for verification. It then constructs an index of media URLs that were actually observed on the page from the DOM and network log. The DOM index includes \texttt{<img src>}, \texttt{<img srcset>}, \texttt{<source srcset>} inside \texttt{<picture>}, \texttt{<video src>}, \texttt{<source src>} inside \texttt{<video>} or \texttt{<audio>}, \texttt{<a href>} entries pointing to media files, CSS \texttt{background-image: url()} entries, and \texttt{<iframe src>} entries for video candidates. URLs observed as image or video responses in the browser network log are also used as source-observation evidence. Each candidate URL is compared against the DOM/network index after URL normalization. BaRA normalizes the scheme, host, query order, and trailing slashes to reduce false mismatches caused by superficial URL differences.

\paragraph{Gate~2 — External accessibility and downloadability.}
BaRA sends an HTTP request to the candidate URL and checks successful retrieval of the actual content. Gate~2 passes when the server returns a successful response and a non-empty response body is obtained. This step filters out URLs that may appear syntactically valid but are inaccessible or not actually downloadable.

\paragraph{Gate~3 — Modality and file-format consistency.}
BaRA verifies whether the Multipurpose Internet Mail Extensions (MIME) type of the downloaded content is consistent with the declared modality of the candidate. It first uses the HTTP \texttt{Content-Type} header. If the header is missing or provides only a generic type, BaRA additionally uses file-signature evidence from the initial bytes of the response body. Image candidates must match an image-family MIME type, whereas video candidates must match a video-family MIME type or a MIME type corresponding to a streaming manifest.

\paragraph{Gate~4 — Site-level non-duplication.}
BaRA computes the SHA-256 hash of the downloaded artifact and compares it against the hash set of artifacts that have already been included from the same site. If the same hash already exists, the candidate is marked as a duplicate and excluded. Deduplication is performed at the site level rather than globally. Therefore, the same image or video appearing on different websites is treated as distinct artifacts.

\paragraph{Gate~5 — Derived hallucination signal.}
Gate~5 is not an independent additional check, but a hallucination signal derived from the results of Gate~1 and Gate~2. If a candidate URL is neither observed in the source DOM or network log nor downloadable, it is marked as having high hallucination risk. Conversely, if either source observation or download validity succeeds, the candidate is considered to have low hallucination risk. This is based on the assumption that a real artifact URL is typically either observed from the source page or directly downloadable.

\paragraph{Final inclusion decision for image/video.}
In strict mode, BaRA includes only image/video artifacts that satisfy source observation, download validity, MIME type consistency, site-level non-duplication, and low hallucination risk. For each candidate, BaRA stores a verification record containing gate-level outcomes, observation evidence, HTTP status, MIME type, file hash, duplicate status, hallucination risk, the final inclusion decision, and exclusion reasons.

\subsubsection{Text Verification}

Unlike image and video artifacts, a text artifact is not an externally downloadable URL object. It is a textual span extracted from the source page. Therefore, BaRA does not apply the URL-level gates above to text artifacts. Instead, it applies two text grounding checks to determine whether a text candidate is grounded in the source-page DOM text.

BaRA first removes non-content elements such as \texttt{<script>}, \texttt{<style>}, \texttt{<noscript>}, and \texttt{<template>} from the rendered DOM, and then extracts the remaining visible text. The DOM text and the candidate text are normalized using the same procedure: HTML entities are decoded, non-breaking spaces are replaced, whitespace runs are collapsed, and the resulting string is lowercased.

\paragraph{Text Gate~1 — Exact DOM substring.}
BaRA first checks whether the normalized candidate text appears as an exact substring of the normalized DOM text. Passing this check means that the text appears verbatim on the source page, and therefore provides the most direct evidence that the text artifact is source-grounded.

\paragraph{Text Gate~2 — Token Jaccard fallback.}
If Text Gate~1 fails, BaRA applies a fallback based on token-level Jaccard similarity. It compares the candidate text against sliding windows over the DOM text and identifies the most similar DOM span. If the Jaccard similarity is above a predefined threshold, the candidate is treated as a source-grounded paraphrase or minor reformatting. In the current implementation, the threshold is set to 0.1. This lenient threshold preserves near-match text that does not satisfy exact substring matching but still has minimal token overlap with the source DOM.

A text candidate that fails both Text Gate~1 and Text Gate~2 is excluded as hallucinated text that is not grounded in the source page. Earlier versions of the design considered boilerplate detection and text deduplication, but these filters were not included in the final verification policy because they risk over-filtering legitimate content such as short titles, repeated category labels, and news section headers.

\paragraph{Final inclusion decision for text.}
A text artifact is included in the final dataset only if it passes either the exact DOM substring check or the token-level Jaccard fallback. Included text is stored together with metadata such as the original text, source page, page index, observation channel, text similarity, and verification timestamp. In the implementation, image/video and text verification results are stored using a shared verification record schema by mapping text grounding outcomes into common record fields. Methodologically, however, text verification is a separate grounding procedure from the URL-level five-gate pipeline used for image and video artifacts.

\section{Prompts}

\label{sec:appendix-prompts}

\subsection{Generalized Bounded-Exploration Prompt}

\label{sec:appendix-generalized-prompts}

For prompt-based baselines and the \textit{w/o BFS} ablation, the Step~1 valid-link discovery prompt implements a generalized bounded-exploration policy at the prompt level rather than BaRA's deterministic DOM-based BFS link-discovery procedure. The agent is restricted to discovering site-internal URLs within the registrable domain of the start URL and its subdomains, which prevents drift toward third-party pages that are not aligned with the data collection objective. The exploration is bounded by a hard \texttt{max\_pages} limit on the number of distinct pages visited. The prompt also specifies normalization and exclusion rules to improve the relevance and quality of the discovered site-internal URL set. For deduplication, URLs differing only in fragments or trailing slashes are treated as identical. For relevance, non-browsable resources are filtered out by both path patterns (\texttt{/privacy}, \texttt{/terms}, \texttt{/login}, \texttt{/signup}) and file extensions, covering archive and binary files (e.g.\ \texttt{.pdf}, \texttt{.zip}, \texttt{.exe}), image files (e.g.\ \texttt{.jpg}, \texttt{.png}, \texttt{.svg}), and video and streaming formats (e.g.\ \texttt{.mp4}, \texttt{.m3u8}, \texttt{.webm}).

The same generalized prompt is used in the \textit{w/o BFS} ablation (Section~\ref{sec:ablation}), where BaRA's deterministic BFS link discovery is replaced with prompt-level exploration; this keeps the prompt-level exploration policy fixed across the relevant comparisons and attributes the ablation effect to replacing BaRA's DOM-based BFS link-discovery control layer with prompt-level exploration, rather than to prompt design differences.

\label{sec:app-generalized-prompt}

\begin{Verbatim}
You collect internal links for the same registrable domain (or its subdomains)
as the start URL.

Exclude (these are not browsable pages — skip them entirely):
  - Paths containing: /privacy, /terms, /login, /signup
  - Archive / binary: .pdf, .zip, .rar, .7z, .apk, .dmg, .exe
  - Image files:      .jpg, .jpeg, .png, .gif, .webp, .svg
  - Video / streaming: .mp4, .mov, .avi, .wmv, .flv, .mkv, .webm, .m4v, .m3u8, .ts

[Start URL] {first_url}
[Limit] max_pages = {max_pages}

[Task]
Browse the site and gather internal URLs. Stop after visiting at most
max_pages distinct pages.

Do not stop at the start page. Follow internal links you find to discover
sub-pages (category pages, paginated lists, deeper navigation, item detail
pages). Keep exploring until you reach max_pages or have exhausted all
reachable same-site pages.

Each URL must appear at most once in the output. Treat URLs as identical
when they differ only in fragments (#...) or trailing slashes.

[Output] JSON ONLY (also save to [general_link.json]):
{{
  "start_url": "{first_url}",
  "max_pages": {max_pages},
  "urls": ["...", "...", ...]
}}
\end{Verbatim}

\subsection{Single-Page Extraction Prompt}

\label{app:single-page-extraction-prompt}

This appendix reports the Step~2 single-page extraction prompt used in the constrained per-page extraction protocol. The agent must remain on the target URL, use bounded scrolling, and return raw text together with candidate image and video URLs in a predefined structured output format. These candidate URLs are then normalized and passed to the downstream rule-based artifact verification pipeline, which checks provenance, accessibility, download validity, modality consistency, and duplication. This design is motivated by the fact that the stability of this stage depends more on the length and parseability of the textual output returned by the agent than on the underlying media files themselves. Consequently, instead of directly downloading images and videos, the prompt requires the agent to list image URLs and video URLs. The extracted media URL candidates are then normalized and verified by dedicated deterministic functions, and only validated media artifacts are downloaded and retained. This separation of responsibilities is consistently maintained throughout the downstream implementation.

\label{sec:app-extraction-prompt}

\begin{Verbatim}
Step 1: Go to {sub_url} and STAY on this specific page.
Step 2: Restricted Extraction (Single Page Only)
- DO NOT click any links, DO NOT navigate to other URLs.
- Mandatory: Scroll to the bottom of the page repeatedly until no new content loads.
Step 3: Comprehensive Data Extraction
- Extract EVERY piece of raw text, image URLs, and video URLs.
- All available image URLs (ending with .jpg, .png, .gif, .jpeg) that appear on that page.
- All available video URLs (ending with .mp4, .mov, .avi, .wmv, .flv, .mkv, .webm, .m4v, .m3u8, .ts) that appear on that page.
Step 4: Strict Output Format for Post-Processing
- Provide the data using the following specific delimiters.
- Do NOT summarize or use '...' to truncate text. List everything found.
- If a category is empty, write 'None' between the tags.

[Page URL]

- Text(s): (Extract all text blocks here. Each paragraph or block must be on a new line. No bullet points or decorative symbols)

- Image(s): (List all full Image URLs here, one per line)

- Video(s): (List all full Video URLs here, one per line)

TERMINATE the task immediately after providing this structured output.
\end{Verbatim}

\subsection{Failure-Analysis Prompt}

\label{app:failure-prompt}

This prompt was designed to operationalize the history-based self-reflection mechanism. Since BaRA recovers from unsuccessful attempts by revising the next instruction rather than terminating immediately, the prompt explicitly provides the original task together with the accumulated execution history and partial outputs and asks the model to identify the failure type, the specific failure point, and concrete prompt-level improvements. The prompt also requires an improved version of the original task so that the reflection step produces an actionable next-step instruction, rather than only a post-hoc explanation of the error. Finally, the response follows a fixed JSON schema, and the reflection-guided retry pipeline uses the diagnosis and revised instruction directly.

\label{sec:app-failure-prompt}

\begin{Verbatim}
You are an expert in analyzing the causes of web scraping task failures and suggesting improvements.

[Original Task]
{original_task}

[Long-term Memory so far (Attempt {attempt_number})]
{json.dumps(long_term_memories, ensure_ascii=False, indent=2)}

[Analysis Request]
Please analyze why the above task failed and provide the following:

1. Failure Analysis:
   - What type of failure it is (e.g., insufficient data, access permissions, technical issues, etc.)
   - The specific point of failure
   - Why this failure occurred

2. Improvement Suggestions:
   - Specific instructions that should be added to the prompt
   - Parts that should be clarified
   - Missing considerations that should be included

3. Improved Prompt:
   - A new version of the original task prompt with improvements
   - Concrete steps that address the identified failure causes

Respond in JSON format:
{{
    "failure_analysis": "Detailed analysis of the failure",
    "failure_type": "Type of failure",
    "improvements": ["Improvement 1", "Improvement 2"],
    "improved_prompt": "Improved task prompt"
}}
\end{Verbatim}

\subsection{Success-Evaluation Prompt}

\label{app:success-prompt}

This prompt was designed as a lightweight and intentionally generous attempt-level success evaluator for the reflection-guided retry pipeline. In BaRA, an attempt is considered useful whenever it yields a usable signal, such as a small number of links, brief but relevant text, or image/video URLs, because the goal is robust collection under fixed interaction budgets rather than all-or-nothing task completion. For this reason, the prompt defines success broadly and marks failure only when no useful data is extracted or the output is clearly irrelevant to the task. Providing both the original task and the final extracted content allows the evaluator to judge whether the produced data is relevant to the intended collection target, as well as whether any data was produced at all. The JSON output further separates the binary decision from its explanation and a short summary of the extracted signal for downstream control and analysis.

\label{sec:app-success-prompt}

\begin{Verbatim}
You are an expert in evaluating the success of web scraping tasks.

[Original Task]
{original_task}

[Last Extracted Content (Final Result)]
{json.dumps(evaluation_data, ensure_ascii=False, indent=2)}

[Evaluation Request]
Please determine whether the above task was successful or not.

Success Criteria (Task is SUCCESSFUL if ANY of the following is true):
1. Links were successfully extracted from the page (even if only a few)
2. Text content was extracted and contains relevant information
3. Images or videos were found and extracted
4. The scraping process completed without major errors
5. The extracted content is related to the original task requirements

IMPORTANT: Be GENEROUS in your evaluation. If the task extracted ANY useful data (links, text, images, videos), consider it a SUCCESS. Only mark as FAILURE if:
- No data was extracted at all
- The extraction process completely failed
- The extracted content is completely irrelevant to the task

Specific Evaluation Points:
- Check if links array contains valid URLs (even if just a few)
- Check if text content was extracted (even if brief)
- Check if images or videos were found
- Consider the task successful if it gathered useful information, even if not perfect

Respond in JSON format:
{{
    "is_successful": true/false,
    "reason": "Detailed reason for success/failure based on what was actually extracted",
    "extracted_data_summary": "Summary of what was actually extracted and why it should be considered success/failure"
}}
\end{Verbatim}

\section{Benchmark Construction Details}

\label{sec:appendix-select website}

\subsection{Synthetic Website Configuration}

\label{sec:appendix-synthetic website configuration}

We built the synthetic corpus by explicitly separating \emph{fixed site structure} from \emph{topic-specific content generation}. For each of the 50 topics in the JSON topic specification, the generator first selected a layout specification from a small family of HTML5 site templates and then filled that structure with topic-conditioned titles, short descriptions, section headers, navigation labels, and captions. Navigational pages use relative internal links, whereas image and video asset URLs remain absolute, explicitly verified external URLs. This separation was useful for benchmark design because it preserves a fixed, inspectable link graph for evaluation while still allowing the site's visible content to vary substantially across topics.

\paragraph{Layout family and depth control.}

The template family was designed to vary the difficulty of the site-internal link discovery without altering the multimodal artifact extraction objective. In the shallowest configuration, all content for a topic is presented in a single \texttt{index.html}. A second configuration uses a small multi-page site in which the root page links to dedicated image and video pages together with one or more text pages (e.g., \texttt{image/image.html}, \texttt{video/video.html}, and \texttt{text\_k.html}). A deeper variant inserts a portal-style entry page at the root and nests the substantive content site under a secondary directory such as \texttt{section\_level\_2/}. These variants preserve comparable topical coverage while systematically varying the number of navigation decisions required before the agent reaches the pages containing the artifact targets. As a result, the synthetic benchmark stresses both direct page-level extraction and more structured multi-hop link discovery under the same evaluation protocol.

\paragraph{Content generation and media handling.}

Topic-specific copy was produced with Google Gemini, which generated short natural-language passages and captions intended to be informative but not excessively repetitive across sites. We intentionally used the model for semantic framing rather than as the final authority on downloadable media URLs. Candidate image and video assets were therefore validated before inclusion, and any broken, unstable, or otherwise unsuitable references were replaced with verified assets retrieved from Unsplash and Pixabay, respectively. In other words, the textual layer is model-assisted, whereas the media layer is grounded by explicit URL validation. We also allowed modest variation in the number of media items and text subpages across topics so that agents could not rely on a single fixed cardinality pattern when collecting artifacts.

\paragraph{Reference recovery and benchmark scope.}

The benchmark reference sets are derived directly from the generated static sites. For download-valid multimodal artifact extraction, we collect the text-bearing pages along with the validated image and video URLs embedded in them, and apply the same normalization rules used during scoring. This design makes the benchmark fully inspectable: once the corpus is generated, repeated runs operate over the same site graph and the same download-valid inventory. We intentionally excluded authentication, form submission, and heavy client-side rendering from this synthetic suite. Those behaviors are covered separately by the real-world benchmark; here, we use a controlled environment that isolates link discovery and page-level extraction errors from nondeterministic website updates.

\begin{table*}[t]
\caption{Cluster labels and representative terms for the selected 50 real-world websites.}
\label{tab:cluster_labels_selected50}
\centering
\begingroup
\small
\setlength{\tabcolsep}{4pt}
\renewcommand{\arraystretch}{1.08}

\begin{tabular}{@{}r p{0.31\textwidth} p{0.45\textwidth} r@{}}
\toprule
\textbf{Cluster ID} & \textbf{Label} & \textbf{Representative terms} & \textbf{Count} \\
\midrule
0  & Restaurants \& dining reviews & restaurant, dining, dishes, reviews, store, time & 4 \\
1  & Research, education \& knowledge & research, library, learning, canada, world & 4 \\
2  & AI \& software industry news & claude, agents, aws, skill, register & 4 \\
3  & General tech \& current affairs (mixed) & self, chat, tech, hosting, attack & 4 \\
4  & Government / public \& online gaming & government, national, cloud, netent, pragmatic & 4 \\
5  & Web development \& design & web, css, webkit, development, design & 5 \\
6  & Open-source software & open, source, distributed, mit, extendable & 4 \\
7  & Travel, lodging \& shopping & hotels, games, printing, buy, offers & 4 \\
8  & Software \& automation & software, automation, automate, offline, data & 5 \\
9  & Personal blogs & addyosmani, fireball, atkinson, focusing, matters & 4 \\
10 & News \& finance & news, financial, market, finance, headlines & 4 \\
11 & Software dev \& databases & databases, rust, legacy, standards, metadata & 4 \\
\bottomrule
\end{tabular}

\endgroup

\end{table*}

\begin{table*}[t]
\caption{
Selected real-world websites used in the benchmark, entries 1--25.
``Src.'' denotes the source from which the site originated, where HN denotes Hacker News.
``Cl.'' denotes the topic cluster ID.
Cluster labels are listed in Table~\ref{tab:cluster_labels_selected50}.
}
\label{tab:final50-sites-a}
\centering
\begingroup
\scriptsize
\urlstyle{same}
\setlength{\tabcolsep}{3pt}
\renewcommand{\arraystretch}{1.05}

\begin{tabularx}{\textwidth}{@{}rZlcc@{}}
\toprule
\textbf{No.} & \textbf{Entry URL} & \textbf{Src.} & \textbf{TLD} & \textbf{Cl.} \\
\midrule
1  & \url{https://blueink.com/} & Tranco & .com & 5 \\
2  & \url{https://sellerassistant.app/} & Tranco & .app & 8 \\
3  & \url{https://smartvsglobe.com/} & Tranco & .com & 9 \\
4  & \url{https://historicacanada.ca/} & Tranco & .ca & 1 \\
5  & \url{https://3v4l.org/} & Tranco & .org & 3 \\
6  & \url{https://royal-lama-casino.com/} & Tranco & .com & 4 \\
7  & \url{https://the-yeatman-hotel.com/} & Tranco & .com & 7 \\
8  & \url{https://itbrief.com.au/} & Tranco & .com.au & 10 \\
9  & \url{https://goto-restaurants.com/} & Tranco & .com & 0 \\
10 & \url{https://jstree.com/} & Tranco & .com & 6 \\
11 & \url{https://www.theregister.com/2025/08/21/aws_ceo_entry_level_jobs_opinion/} & HN & .com & 2 \\
12 & \url{https://nesbitt.io/2025/12/26/how-uv-got-so-fast.html} & HN & .io & 11 \\
13 & \url{https://webzoneexpertz.com.au/} & Tranco & .com.au & 5 \\
14 & \url{https://fyidocs.com/} & Tranco & .com & 8 \\
15 & \url{https://altavillaspa.com/} & Tranco & .com & 9 \\
16 & \url{https://enclose.horse/} & HN & .horse & 1 \\
17 & \url{https://pase.com.mx/} & Tranco & .com.mx & 3 \\
18 & \url{https://littlechapel.com/} & Tranco & .com & 4 \\
19 & \url{https://balenchy.in/} & Tranco & .in & 7 \\
20 & \url{https://bombaytimes.com/} & Tranco & .com & 10 \\
21 & \url{https://pageplay.com/} & Tranco & .com & 0 \\
22 & \url{https://mahadk.com/posts/slack} & HN & .com & 6 \\
23 & \url{https://pilk.website/3/facebook-is-absolutely-cooked} & HN & .website & 2 \\
24 & \url{https://www.theocharis.dev/blog/kidnapped-by-deutsche-bahn/} & HN & .dev & 11 \\
25 & \url{https://hey.paris/posts/appleid/} & HN & .paris & 5 \\
\bottomrule
\end{tabularx}

\endgroup

\end{table*}

\begin{table*}[t]
\caption{
Selected real-world websites used in the benchmark, entries 26--50.
``Src.'' denotes the source from which the site originated, where HN denotes Hacker News.
``Cl.'' denotes the topic cluster ID.
Cluster labels are listed in Table~\ref{tab:cluster_labels_selected50}.
}
\label{tab:final50-sites-b}
\centering
\begingroup
\scriptsize
\urlstyle{same}
\setlength{\tabcolsep}{3pt}
\renewcommand{\arraystretch}{1.05}

\begin{tabularx}{\textwidth}{@{}rZlcc@{}}
\toprule
\textbf{No.} & \textbf{Entry URL} & \textbf{Src.} & \textbf{TLD} & \textbf{Cl.} \\
\midrule
26 & \url{https://www.eff.org/deeplinks/2026/04/google-broke-its-promise-me-now-ice-has-my-data} & HN & .org & 8 \\
27 & \url{https://addyosmani.com/blog/21-lessons/} & HN & .com & 9 \\
28 & \url{https://www.researchsquare.com/article/rs-6079807/v1} & HN & .com & 1 \\
29 & \url{https://pokecardgenerator.com/} & Tranco & .com & 3 \\
30 & \url{https://nationalpopularvote.com/} & Tranco & .com & 4 \\
31 & \url{https://heigame.xyz/} & Tranco & .xyz & 7 \\
32 & \url{https://unofficialroyalty.com/} & Tranco & .com & 10 \\
33 & \url{https://evanhahn.com/scripts-i-wrote-that-i-use-all-the-time/} & HN & .com & 0 \\
34 & \url{https://piechowski.io/post/git-commands-before-reading-code/} & HN & .io & 6 \\
35 & \url{https://manualdousuario.net/en/mozilla-firefox-window-ai/} & HN & .net & 2 \\
36 & \url{https://turso.tech/blog/working-on-databases-from-prison} & HN & .tech & 11 \\
37 & \url{https://plainvanillaweb.com/index.html} & HN & .com & 5 \\
38 & \url{https://ahooy.pl/} & Tranco & .pl & 8 \\
39 & \url{https://daringfireball.net/linked/2025/06/07/bill-atkinson-rip} & HN & .net & 9 \\
40 & \url{https://benholmen.com/blog/kilopixel/} & HN & .com & 1 \\
41 & \url{https://vlab.co.in/} & Tranco & .co.in & 3 \\
42 & \url{https://wiley.law/} & Tranco & .law & 4 \\
43 & \url{https://passforless.com/} & Tranco & .com & 7 \\
44 & \url{https://news.ycombinator.com/newsguidelines.html} & HN & .com & 10 \\
45 & \url{https://sethpurcell.com/writing/screens-in-museums/} & HN & .com & 0 \\
46 & \url{https://openai.com/open-models/} & HN & .com & 6 \\
47 & \url{https://www.geoffreylitt.com/2025/07/27/enough-ai-copilots-we-need-ai-huds} & HN & .com & 2 \\
48 & \url{https://www.cs.cmu.edu/~pavlo/blog/2026/01/2025-databases-retrospective.html} & HN & .edu & 11 \\
49 & \url{https://tenderlovemaking.com/2025/09/17/apple-photos-app-corrupts-images/} & HN & .com & 5 \\
50 & \url{https://lawsofsoftwareengineering.com} & HN & .com & 8 \\
\bottomrule
\end{tabularx}

\endgroup

\end{table*}

\subsection{Reference Construction for the Synthetic Benchmark}
\label{sec:synth-reference}
Step~1 valid-link discovery evaluation does not require an annotated link inventory or reference crawl: as defined in Section~\ref{sec:metrics}, it measures the fraction of discovered URLs that pass two-stage liveness verification using an HTTP request followed by a Playwright-based accessibility check. Reference construction is therefore needed only for Step~2 multimodal artifact extraction evaluation.
For each site in the synthetic corpus, we enumerate the HTML pages under its directory and map them to their hosted page URLs. Each page is parsed into three per-page reference sets used by the metrics in Section~\ref{sec:metrics}: a word set for the on-page text, and URL sets for the referenced images and videos.
At scoring time, system outputs are matched to references at the page granularity by canonical URL rather than visited-page order. The URL and text normalization rules applied on both sides, together with the page-to-site-to-corpus macro averaging and the treatment of modalities that are empty in both prediction and reference, are described in Appendix~\ref{sec:app-metrics}.
Because the synthetic corpus is static, this construction is fully deterministic and requires no manual annotation.

\subsection{Real-World Benchmark Website Selection}
\label{sec:appendix-realworld-selection}

To complement the controlled synthetic benchmark with live web conditions, we construct the real-world benchmark as a set of 50 English public websites, no two sharing the same registrable domain (eTLD+1). The selection deliberately avoids hand-curated lists in favor of a reproducible pipeline, so that the resulting benchmark reflects the topical and structural diversity of heterogeneous live websites, including changing page layouts, varied content structures, and diverse interaction patterns, rather than the topical preferences of an annotator.

Candidate websites are sampled from two independent sources covering complementary strata of the web: Tranco long-tail domains sampled from ranks 200k--800k, and Hacker News stories of at least 20 points within the past 12 months (via the Algolia API). Every candidate website passes through a four-stage accessibility and content filtering pipeline: (i) \texttt{robots.txt} compliance under our identifying user-agent and rejection of authentication- or subscription-wall-suggestive paths (\texttt{login}, \texttt{signup}, \texttt{account}, \texttt{register}, \texttt{paywall}, \texttt{members}, \texttt{premium}); (ii) HTTP probing requiring 200~OK, \texttt{text/html}, body $\le 5$\,MB, and the absence of bot-challenge signatures (Cloudflare interstitial, Akamai access-denied, generic JavaScript-cookie walls); (iii) language and structural checks requiring at least 300 characters of script-stripped text, an English language tag, and \texttt{langdetect} confidence $\ge 0.85$, with explicit rejection of password-protected pages, known paywall or subscription-wall eTLD+1s (NYT, WSJ, FT, Bloomberg), and single-page-application shells; (iv) a depth-2 site-internal Playwright traversal, restricted to the same registrable domain and subdomains, verifying that at least one image or video candidate, such as an \texttt{<img>} or \texttt{<video>} element, is discoverable within two link hops of the entry URL. The final gate verifies discoverability of media within two hops, rather than the broader claim that a media artifact exists somewhere on the website.

The final 50 websites are produced by a topic-balanced selection procedure with diversity constraints over the admitted candidates. We compute TF-IDF vectors (2{,}000 features, English stop words, unigrams and bigrams) over each candidate's \texttt{<title>} and meta description, cluster candidates into $K = 12$ topic clusters with KMeans (\texttt{random\_state}=13, \texttt{n\_init}=10), and pick round-robin from clusters in current-smallest order. Selection treats eTLD+1 uniqueness as a hard constraint and applies soft targets discouraging source, TLD, and topic-cluster concentration; soft targets are progressively relaxed when satisfying all of them simultaneously is infeasible for the admitted candidate set, while eTLD+1 uniqueness is preserved throughout. The final benchmark contains 26 Tranco-originated websites and 24 Hacker-News-originated websites, spans 19 TLDs, and assigns 4--5 websites to each topic cluster. The resulting cluster taxonomy and representative terms are listed in Table~\ref{tab:cluster_labels_selected50}, and the per-site composition (entry URL, source, TLD, assigned cluster) in Tables~\ref{tab:final50-sites-a} and ~\ref{tab:final50-sites-b}. All network operations throughout sourcing and filtering use an identifying user-agent (\texttt{BaRA-Agent-Research/0.1} with a reachable contact email), $\le 1$ request per second globally, $\ge 5$ seconds between requests to the same domain, and immediate candidate-domain exclusion on any \texttt{4xx}/\texttt{5xx} response or \texttt{Retry-After} header without retry.

\begin{table*}[t]
\caption{Runtime configuration used across experiments.}

\label{tab:shared-runtime-config}
\centering

\footnotesize

\begin{tabularx}{\linewidth}{lX}

\toprule

\textbf{Parameter} & \textbf{Value} \\

\midrule

Headless mode & True \\

Max actions per step & 1 \\

LLM API provider & OpenRouter\citep{openrouter_api} \\

Base model name & gemini-3-flash-preview \\

Max (retry) attempts per stage & 2 \\

Random seed & 42 \\

Step~2 temperature & 0.6 \\

Step~2 top-p & 0.7 \\

Step~2 LLM timeout & 240\,s \\

Step~2 step timeout & 360\,s \\

Step~2 total timeout & 900\,s \\

Step~2 supplemental retry total timeout & 1500\,s \\
Verification token-level Jaccard similarity threshold & 0.1 \\
Firecrawl \citep{firecrawl_repo} integration & Enabled only when an API key is provided \\

API Keys & Enabled only when an API key is provided \\

\bottomrule

\end{tabularx}

\end{table*}

\subsection{Reference Construction for the Real-World Benchmark}
\label{sec:appendix-realworld-reference}

Real-world websites are live and time-varying, so their Step~2 reference artifact sets are not constructed ahead of evaluation using the same deterministic procedure as in the synthetic benchmark. We instead build a \emph{post hoc reference set} per site that is shared across all compared methods and never exposed to any method during browsing, extraction, or method-side artifact verification; it is used only for scoring.

Step~1 valid-link discovery evaluation does not require an annotated link inventory, by the same argument as in the synthetic setting: it measures the fraction of agent-discovered URLs that pass two-stage HTTP and Playwright liveness verification, computed directly from each system's output. Reference construction is therefore needed only for Step~2 multimodal artifact extraction.

For Step~2, each page in the reference set is rendered and parsed once into three per-page reference artifact sets used by the metrics in Section~\ref{sec:metrics}: a word set for the on-page text, and URL sets for the referenced images and videos. The set captures each site's state at a fixed timestamp recorded in the released manifest, and all methods are scored against this single set. URL normalization, text normalization, page-level matching by canonical URL, and page-to-site-to-corpus macro averaging follow the rules described in Appendix~\ref{sec:app-metrics}. These rules are identical to those used in the synthetic setting, so that the only deliberate difference between the two benchmarks is the source of the evaluated pages, not the way scores are computed.

\section{Experimental Settings}

\label{app:experiment-set}

\subsection{Runtime Configuration}

\label{app:runtime-configuration}

All browser-agent experiments were run in a headless Chromium environment built on Browser-use. The shared runtime configuration is summarized in Table~\ref{tab:shared-runtime-config}. Following the main experimental setup, we used \texttt{gemini-3-flash-preview} as the backbone model for Step~2 multimodal artifact extraction, where repeated browser-agent execution makes throughput and response latency practically important.

\begin{table*}[t]
\caption{Stage-wise summary of the retry-and-merge rules used in BaRA under fixed interaction budgets.}
\label{tab:retry-and-merge-summary}
\centering
\small
\setlength{\tabcolsep}{5pt}
\renewcommand{\arraystretch}{1.16}
\begin{tabularx}{\textwidth}{>{\RaggedRight\arraybackslash}p{3.15cm}@{\hspace{0.8em}}Y}
\toprule
Stage & Rule summary \\
\midrule
\parbox[t]{\linewidth}{Step~2\\(page-level extraction)}
&
When an attempt fails or returns incomplete output, the next attempt's instruction is revised from execution history and partial output via $p_{t+1}\!\leftarrow\!R(p_t,h_t,y_t)$, and the attempts are merged. Text bullet items are set-deduplicated and emitted in sorted order; image URLs are normalized (query and fragment stripped) and deduplicated with the longest variant retained; video URLs are trimmed and set-deduplicated. \\
\bottomrule
\end{tabularx}

\end{table*}

\subsection{Retry-and-Merge Rules}
\label{app:retry-and-merge-rules}

Table~\ref{tab:retry-and-merge-summary} summarizes BaRA's retry-and-merge rules for Step~2 multimodal artifact extraction under the fixed interaction budget. We focus on Step~2 because retry-and-merge is applied during page-level extraction; Step~1 valid-link discovery is implemented as a single deterministic DOM-based BFS pass with liveness verification, as described in Section~\ref{sec:bfs}.

At Step~2, BaRA uses history-based self-reflection to revise the next prompt from the agent's execution history and the partial output observed so far, in line with Section~\ref{sec:reliability}. Operationally, this mechanism is realized as up to $A$ reflection-based retry attempts on the same page, and candidate outputs are merged across attempts before rule-based artifact verification. When multiple attempts are available for a page, their outputs are aggregated modality-wise. Text artifacts are collected as text blocks after trimming empty lines and filtering \texttt{None}-style placeholders, and the resulting set is emitted in sorted order. Image URLs are normalized for the merge stage by stripping query strings and fragments and then deduplicated by canonical key, with the longest original variant retained per key. Video URLs are trimmed of surrounding whitespace and list markers and then set-deduplicated. Unlike image URLs, video URLs are not stripped of query strings at this pre-verification merge stage, because video sources frequently carry session or playback parameters that distinguish playable from non-playable variants.

The merged candidate set is then passed to rule-based artifact verification (Section~\ref{sec:reliability}), which applies provenance and accessibility checks and retains only artifacts satisfying the modality-specific inclusion criteria, including download validity for media artifacts. This design is intentionally recall-oriented: complementary attempts expose additional lazy-loaded content or recover omitted artifacts in some cases.

Schema-nonconforming outputs, outputs with no usable signal, and outputs that are clearly irrelevant to the target page are handled conservatively. Such outputs are treated as failed or incomplete attempts and trigger a reflection-revised retry when the remaining retry budget allows. At the same time, partially useful extractions are not discarded unnecessarily: if an attempt is missing evidence in one modality, follow-up attempts may still contribute to the merged candidate set for that page, with final correctness determined only after normalization and rule-based artifact verification.




\section{Qualitative Case Study}
\label{sec:qualitative}

Figure~\ref{fig:qualitative-case} provides a qualitative case study on a synthetic cat website, illustrating the two failure modes captured by our evaluation. In Step~1, the task is not to match a precomputed BFS reference tree, but to return URLs that are actually accessible after two-stage liveness verification. Pure LLM and SeeAct predict plausible first-level pages such as gallery or video pages that are not accessible on the site. In contrast, Browser-use and BaRA recover the accessible first-level pages exposed by the website, showing the importance of grounding link discovery in the actual site structure rather than in semantically plausible URL patterns.

The bottom panels illustrate provenance-grounded artifact extraction on the image and video pages. Pure LLM and SeeAct return hallucinated media URLs that look plausible but are not the true page-grounded sources. Browser-use reaches the relevant pages more reliably, but still misses the image artifact and duplicates the video artifact, showing that valid-link discovery alone does not guarantee download-valid multimodal extraction. BaRA recovers the correct Unsplash image source and Pixabay MP4 source without hallucinated or duplicated outputs. This example illustrates how BaRA's liveness verification and artifact verification jointly improve live web data collection: the system must both reach accessible pages and return verifiable artifacts in an accessible form.

\end{document}